\documentclass[journal]{IEEEtran}
\usepackage{graphicx}
\usepackage{amsmath}
\usepackage{subfigure}
\usepackage{rotating}
\usepackage{multirow}
\usepackage{array}
\usepackage{algorithm}
\usepackage{algorithmic}

\begin{document}
%
\title{Noise Filtering, Channel Modeling and Energy Utilization in Wireless Body Area Networks}

\author{\IEEEauthorblockN{B. Manzoor, N. Javaid, A. Bibi, $^{\$}$Z. A. Khan, M. Tahir\\}

        Department of Electrical Engineering, COMSATS\\ Institute of
        Information Technology, Islamabad, Pakistan. \\
        $^{\$}$Faculty of Engineering, Dalhousie University, Halifax, Canada.
        }

\maketitle

\begin{abstract}

Constant monitoring of patients without disturbing their daily activities can be achieved through mobile networks. Sensor nodes distributed in a home environment to provide home assistance gives concept of Wireless Wearable Body Area Networks. Gathering useful information and its transmission to the required destination may face several problems. In this paper we figure out different issues and discuss their possible solutions in order to obtain an optimized infrastructure for the care of elderly people. Different channel models along with their characteristics, noise filtering in different equalization techniques, energy consumption and effect of different impairments have been discussed in our paper. The novelty of this work is that we highlighted multiple issues along with their possible solutions that a BAN infrastructure is still facing.
\end{abstract}


\section{Introduction}

Global trends in growth of elderly people attracts researchers to formulate wireless health monitoring systems. The system use sensors for
continuous monitoring of patients with less interaction to doctors. Wearable Wireless Body Area Networks (WWBAN) gained a tremendous attention in providing such facilities to keep in view the elderly age's, as well as giving proper weight to the economical issues. Moreover, they are providing a reasonable platform to the caregivers in recognition of patterns of diseases through their databases.


The data acquired by the central device is transmitted to the medical servers or to the care givers through wireless network. This communication
have many issues to be addressed. 1) Channel models, 2) effect of different interferences/disturbances, 3) noise filtering and 4) energy consumption are some of the issues which have been addressed separately and there are a number of possible solutions are proposed for these issues.

Proper estimation and equalization of transmitted signal is a key concern in wireless communication. For this purpose, different channel models are used which describes different issues like; path loss, amplitude distortion, clustering and inter arrival-time characteristics for proper
estimation of the signal. Signal can be degraded by Additive White Gaussian Noise (AWGN). Therefore noise filtering is also required in different
equalization schemes for better performance.

Simultaneous communication of sensors yields in different types of interferences and disturbances (ISI, MUI, and NOISE) which affect strongly the
performance of received signal. These disturbances emerge to be a great hurdle in signal reception and can destroy partially or fully the information being sent to the destination. Mean while, retransmission of signals for proper data acquisition can utilize more energy which is not suitable in any case. Energy utilization in wireless BAN is a critical issue due to its infrastructure.


\section{Related Work and Motivation}
Rapid advancements in technology devices many possibilities to address routine issues and hence making life as simpler and cureable as it could be.
WWBAN introduces a unique solution for the cure of elderly peoples throughout the world by continuous monitoring of the patients anywhere and anytime . Although it's a broad domain. However, there is a lot, to be discussed and optimized for making this approach more effective and goal oriented.

In WWBAN different sensors are working simultaneously to attain information to transmit it to the desired destination. Energy utilization in wireless transmission are the main issues which should be given equal importance to extend the life of sensors. In order to extract useful information in the presence of different interferences e.g. Inter Symbol Interferences (ISI), Multi User Interference (MUI), and noise certain algorithms have been proposed. MUD receivers provide an appropriate platform to mitigate the effect of MUI caused by wireless nodes which are communicating asynchronously [1]. Thus, energy utilization for retransmission is optimized.

\begin{figure*}[!t]
\centering
  \includegraphics[width=17cm, height=8cm]{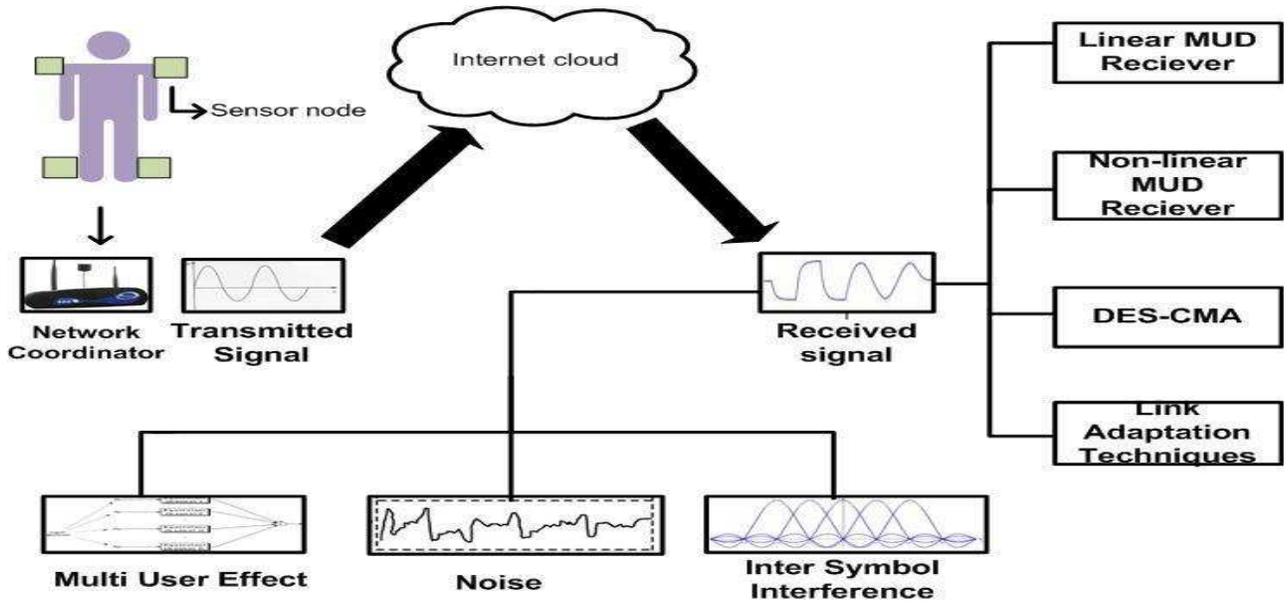}
  \caption{Issues and Possible Solutions in WWBAN Infrastructure}
\end{figure*}

Improved performance of the system by allowing nodes to use Medium Access Control protocol (MAC) among different modulation schemes gave birth to the concept of novel Link Adaptation (LA) strategy [2]. By considering the effects like large SNR and low SIR data rates can be adapted dynamically by reducing the time required for transmission and consequently the interference caused by it [2].

Full bandwidth utilization of the channel without transmission of the training signal presents an idea of blind equalization scheme. Constant Modulus Algorithm (CMA) emerges to be a suitable blind wireless channel equalizer. The equalizer is used to remove  ISI which is produced by dispersive channels [3]. Further studies states that use of dither input signal to the equalizer by a distributed signal before a sign operation being applied can be more goal oriented.

Fig 1 defines the hierarchy of the Wireless Body Area Networks (WBANs), its related issues and some of the possible solutions. The information obtained from the sensor nodes is transmitted to the  personal digital assistant (PDA). This information is then transmitted to the care givers or medical server for continuous monitoring of the patients through internet for further processing. The transmission and reception of this information through any medium may find certain problems like MUI, ISI and noise which can be resolved through different techniques like by using MUD receivers, DES-CMA and LA techniques. In our work we present an overview of different issues which a BAN infrastructure still have to deal with. We highlighted those issues along with some of the possible solutions.

\section{Noise Filtering}
This section describes noise filtering in different equalization techniques for WBAN. We discuss and compare noise
filtering in three different selected scenarios; 1) MUD recievers 2) Novel Link Adaptation, 3) Blind Equalization.

\subsection{Noise Filtering In Multiuser Detection Based Recievers}
Signal are commonly effected by InterSymbol  Interference (ISI) and Additive White Gaussian Noise (AWGN). ISI is minimized through equalization, whereas, noise is still a problem [1].

\vspace{-0.5cm}
\begin{eqnarray}
r=s^*h+n
 \end{eqnarray}
where, $r$=recived signal, $s$=transmitted signal, $h$=channel response and $n$=noise.
Equalizer parameters can be calculated using Wiener-hopf equation, which is presented in [1] as:

\begin{eqnarray}
w=\Upsilon_{rr}\Gamma^-1_{rr}\
 \end{eqnarray}

There is a tradeoff between ISI and noise. Two types of receivers are defined in [1];

A) {Linear Multi User Detector based MUD receivers}

B) {Non-linear Multi User Detector based MUD receivers}

\subsubsection{Linear MUD Receivers}
In MUD based receivers, there is only fixed equalizer parameters [1]. Therefore, these are not suitable for fast varying channels. By implementing Minimum Mean Square Error (MMSE), ISI can be compensated. However, effect of noise is still there. Linear MUD tries to mitigate the effect of MUI and ISI. Whereas, it does not distinguish between noise and MUI. Therefore, another approach named as non-linear MUD is introduced.

\subsubsection{Non-linear MUD Receivers}
In such MUD based receivers there is no fixed equalizer parameter. This is because DFE is used to adjust equalizer parameter according to detected signal quality. Hence, it is more suitable for fast varying channels. Noise becomes less problem when someone try to completely remove ISI (zero forcing) that enhances noise, effect of noise is catered by implementing Minimum Mean Square Error (MMSE) criteria through Non Linear MUD Receivers.

\subsection{Noise Filtering In Link Adaptation For IEEE 802.15.4}
As, it is  discussed in previous section that noise filtering is a problem in interference environment. Link Adaptation is a technique which is used in [2] to minimize the ISI. The basic idea behind this technique is discussed in [2] which is based on, the selection of modulation technique according to channel quality. However, by increasing order of modulation high data rates can be achieved. In wireless environment, degradation is caused due to multi-path. Channel quality depends upon Signal to Noise Ratio (SNR) and Signal to Interference Ratio (SIR). If SNR and SIR is low then channel is not good. Therefore, lower order modulation is chosen, however, this selection leads to increase ISI. If SNR is below some threshold then lower order modulation is chosen which causes decrease in data rate. Otherwise, higher order modulation is selected which causes increase in data rate. If S/N is above some threshold and Packet Delivery (P.F) is below threshold then higher order modulation is avoided.
\begin{figure}[!t]
  \centering
  \subfigure[]{\includegraphics[height=7 cm,width=8 cm]{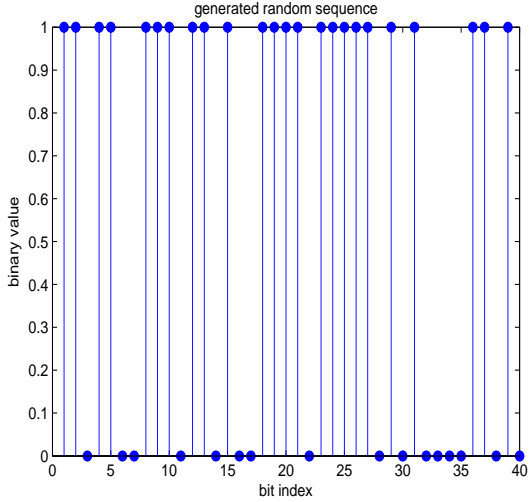}}
  \subfigure[]{\includegraphics[height=6  cm,width=7 cm]{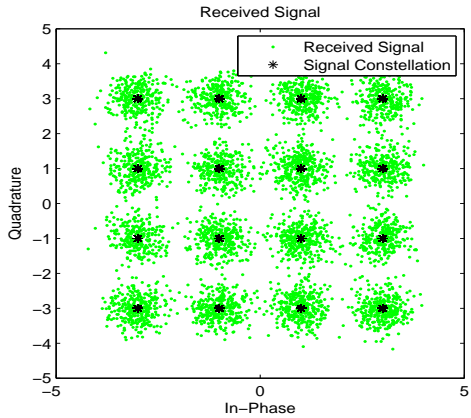}}
  \subfigure[]{\includegraphics[height=6  cm,width=7 cm]{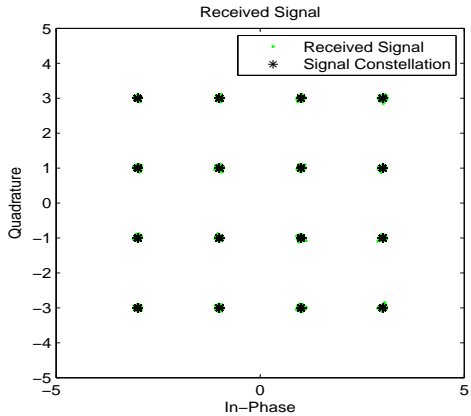}}
  \caption{Performance Evaluation of Channel Models in the Presence of AWGN }
\end{figure}

We compare performance evaluation of channel models in the presence of AWGN. For this purpose Quadrature Amplitude Modulation modulation (16-QAM)  scheme along with AWGN channel for the transmission of data stream is used. The general idea is to change modulation or coding schemes according to response of  channel. The response can be addressed through SNR. The channel is considered 'bad' when SNR is low, due to path loss, shadowing, or fading problems. Whereas considered 'good' when SNR is high. The ratio of bit energy to noise power spectral density is set to 10dB and is then converted to corresponding SNR. A random bit is generated, as shown in fig 2a and then converted in to random symbols that have to be
transmitted through AWGN channel.

The system Bit Error Rate (BER) is computed using 16-QAM modulation scheme, as shown in fig 2b. Moreover, the effect of changing S/N on the quality of received signal is shown in fig 2c.

\subsection{Noise Filtering in Blind Equalization}
For equalization of biological signals using adaptive Blind Equalization Algorithm called as Constant Modulus Algorithm (CMA) [3] is not suitable for noise filtering. Because, training sequence is not sent with information. Therefore, FEC cannot be implemented. All the effect of multipath and noise is on actual information. The detection of signal is based on stochastic information of signal that is stored at receiver. In case of blind equalization, if receiver does not know about channel response then noise filtering becomes difficult. Setting equalizer parameters periodically according to errors received in previous is discussed in [3] minimizes noise. If we go towards zero forcing solution than it leads to enhance noise.
Adaptive algorithm is defined in [3] as:

\begin{eqnarray}
f(n+1)=f(n)+\mu{r(n)}\psi_{dse-cma}(y(n))
 \end{eqnarray}
 where, $f(n+1)$=next equalizer parameter, $f(n)$=current equalizer parameter, $\mu$=step size and $r(n)$=received signal.

\section{Channel Modeling}
In this section, we discuss different channel models and characteristics of these models for wireless communication in different standards of WBAN. These models are used for estimation and equalization of transmitted signals. We compare and discuss three channel models for WBAN.

\subsection{Channel Model for IEEE 802.15.4a}
Channel model for MUD based receivers is recently proposed for IEEE 802.15.4a is given in [1]. One of the reasons for ISI is channel memory. Traditional receiver (DFE, MMSE, MLSE) knows about channel characteristics. Analysis shows that electromagnetic propagation at UWB range (3.1-10.6) GHz is negligible through human body. Transmitted signals reaches at receiver in two different ways: 1: Around human body, 2: Through reflection from surrounding. Therefore, WBAN channel posseses significantly different path loss, amplitude distortion, clustering and inter arrival-time characteristics. We summarize simulation results for WBAN's channel model as: (A) For an outdoor channel (Outdoor-BAN), measurements [1] indicate that there are always two clusters of multipath components due to the initial wave diffracting around the body, and reflections from the ground. Therefore, number of clusters is always two and does not need to be defined as a stochastic process as defined for other scenarios. Moreover, depending upon exact position of transmitter on human body intercluster arrival time distribution is deterministic. (B) A human wearing WBAN devices in an indoor environment is defined as an Indoor channel. Two deterministic clusters due to stochastic nature of multipath (reflections from the indoor environment) should be taken into account, as discussed in [1].
Channel model corresponding to the propagation for outdoor environment [1]:

\begin{eqnarray}
h_{body}(t)=\sum_{k=0}^\infty\beta_kexp(j\phi_k)\delta(t-\triangle{k})
\end{eqnarray}

where, ${\beta_k}$  is the $k^{th}$ element of vector Y,  ${\phi_k}$ is uniformly distributed random variable and ${\triangle}$ is size of bin.

Model parameters are taken  for 'front', 'side' and 'back' positions of receiver on human body. Reflections from ground introduces a delay $\tau_{ground}$ in channel model and thus, overall channel model is given as :

\begin{eqnarray}
h_{ground}(t)=\sum_{k=0}^\infty\beta_kexp(j\phi_k)\delta(t-\triangle{k}-\tau_{ground})
\end{eqnarray}

By assuming that reflections from ground are uncorrelated with components diffracting around body, then complete channel response and is
represented in [1] as:

\begin{eqnarray}
h_{outdoor-BAN}(t)=h_{body}(t)+h_{ground}(t)
\end{eqnarray}

For an Indoor-BAN channel, reflections from surroundings create multipath. Multipath distribution is stochastic in nature therefore, clusters are modeled using Poisson or Weibull distributed random variables. Whereas, intra-cluster distribution is dense, so multipath arrivals within a cluster are assumed to be uniform [1]. The channel model due to reflections from surroundings is given in [1] as:
\begin{eqnarray}
h_{ref}(t)=\frac{X}{\sqrt{E}}\sum_{l=0}^\infty\sum_{K=0}^\infty\beta_l,_Kexp(j\phi_l,k)\delta(t-\triangle([\tau_l/\Delta]+k))
\end{eqnarray}

\begin{eqnarray}
20log_{10}(\beta_l,k)=\Gamma_{\tau_l}+\Upsilon(\tau_l+\Delta_k)+\sigma_{\Gamma}n_l+\sigma_{\Upsilon}n_k
\end{eqnarray}

\begin{eqnarray}
p(\tau_l/\tau_l-1)=\frac{1}{\beta} e^-(\tau_l-\tau_l-1/\beta)
\end{eqnarray}

where, $\beta_l$, $K$ represents the amplitude in $k^{th}$ bin and in $l^{th}$ cluster modeled in [1], $\tau_l$ is the cluster arrival time modeled using a Poisson distribution, $\beta$ represent the mean arrival rate.

Average energy of the clusters decays exponentially at a rate of $\Gamma dB/ns$ and the terms within the cluster decay at the rate of $\gamma dB/ns$. Lognormal fading in the clusters and within the clusters is modeled using $\sigma\Gamma$ and $\sigma\gamma$, respectively. Here, $n_l$ and $n_k$ represent uncorrelated normal random variables with unit mean and unit variance.
$\phi_l$, $k$ are uniformly distributed random variables and $\Delta$ is bin size [1].

If we assume that reflections from indoor environment and ground are uncorrelated with components diffracting around the body, then complete channel response from [1] is:

\begin{eqnarray}
h_{indoor-BAN}(t)=h_{body}(t)+h_{ground}(t)+h_{ref}(t)
\end{eqnarray}

\subsection{Channel Model for IEEE 802.15.4}
Channel model that is used for a proposed novel Link Adaptation (LA) strategy [2], where nodes select their modulation schemes according to the
experienced channel quality and level of interference. The general idea is to change modulation and/or coding scheme in such a way that it lowers bit rates (more robust modulation/coding when the overall bandwidth is set) in case of "bad" channel conditions. Whereas, high bit rates when the channel is "good". The bad channel signifies low SNR, due to path loss, shadowing or fading problems, or when SIR is low. In both cases, the use of a more robust modulation/coding is supposed to reduce probability of error. The channel loss, denoted as $A$, is modeled according to 3GPP proposal for WBANs, as discussed in [2]:

\begin{eqnarray}
A(d)[dB]=A_0+10n\log(d/d_0)+s
\end{eqnarray}
where $d$ is the distance between transmitter and receiver, $A_0$ defines path loss in $dB$ at a reference distance, $d_0$, $n$ specifies path loss exponent and $s$ represents a random Gaussian variable having zero mean and standard deviation $\sigma$.

\begin{algorithm}[H]
\caption{Novel Link Adaptation}
\begin{algorithmic}[1]
\STATE $Set \;of \;nodes \;in \;BAN \leftarrow \;N$
\STATE $Received \;beacon \;power\leftarrow \;B_{RP}$
\STATE $Failure \;Probability \;of \;received \;packet \leftarrow \;P_F$
\STATE $Threshold \;for \;failure \;probability \;pcket \leftarrow \;TH_{PF}$
\STATE $Bit \;rate \leftarrow \;B_R$
\STATE $Number \;of \;sensor \;nodes \;in \;BAN \leftarrow \;n$
 \FORALL {$n\in N$}
  \STATE $Find \;B_{RP}  \;for \;every \;n ;B_{RP}(n)$
  \STATE $Set \;B_{RP}(n)\leftarrow \;P_{n}$
 \IF{$\;P_{n}- $\;SNR $\;<$ $\;TH_{SNR}$}
\STATE {$B_{R}= B_{R}^{--}$}
 \ENDIF
\IF{$\;P_F$ $\;>$ $\;TH_{PF}$}
\STATE {$B_{R}= B_{R}^{++}$}
\ELSE
\STATE {$B_{R}= B_{R}$}
 \ENDIF
  \ENDFOR
 \end{algorithmic}
 \end{algorithm}


\subsection{Geometrical Based Hyperbolically Distributed Scatters channel model}
 Geometrical-Based Hyperbolically Distributed Scatterers (GBHDS) channel model is discussed in [3], for macro cell environments is considered in this section. A comprehensive study of this model proved to be more realistic than other models, as tested against practical data [3]. GBHDS for macro cell environment channel model assumes that scatterers are arranged within a circle of radius R around mobile. The distances $r$ between Mobile Station (MS) and scatterers are distributed according to hyperbolic Probability Density Function (pdf). The Geometrical Scatterers Density Function (GSDF) for this model, $fr_(r)$, is given in [3] as:

\begin{eqnarray}
f_r(r)=\{\frac{a}{\tanh(ar)\cosh^2(ar)}
\end{eqnarray}
where, $R$ is the radius of the circle enclosing the scatterers, and applicable values of $\alpha$ lie in interval (0, 1) [3]. GBHDS results show that there is a good match between model result and measurement data which is reported for Outdoor environment. Fig 4 shows a comparison of the results for the Direction of Arrival (DOA) pdf for the GBHDS channel model, DOA pdf for the Geometrical Based Single Bounce Macro Cell (GBSBM) channel model and Gaussian Scatterer's Density (GSD) channel model. From fig 4 it is clear that there is a good match between the GBHDS channel model result and measurement data. Whereas, GSD model fails in the DOAs close to Line-of-Sight (LOS).

\begin{figure}[ht]
  \centering
  \subfigure[]{\includegraphics[scale=0.5]{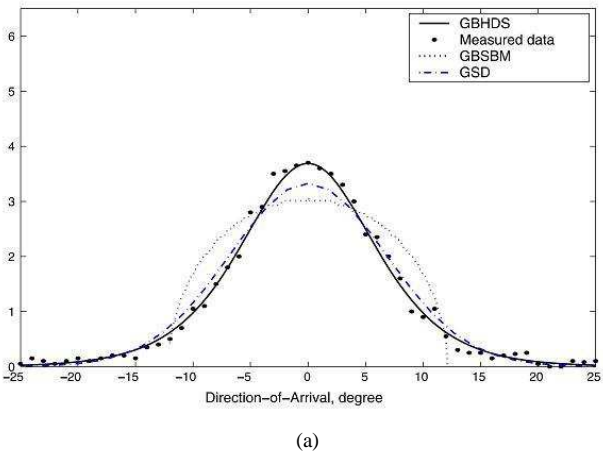}}
  \caption{D.O.A Comparison for GBHDS, GBSBM AND GSD Channel Models}
\end{figure}

\begin{figure}[ht]
  \centering
  \includegraphics[height=7.0  cm,width=10 cm]{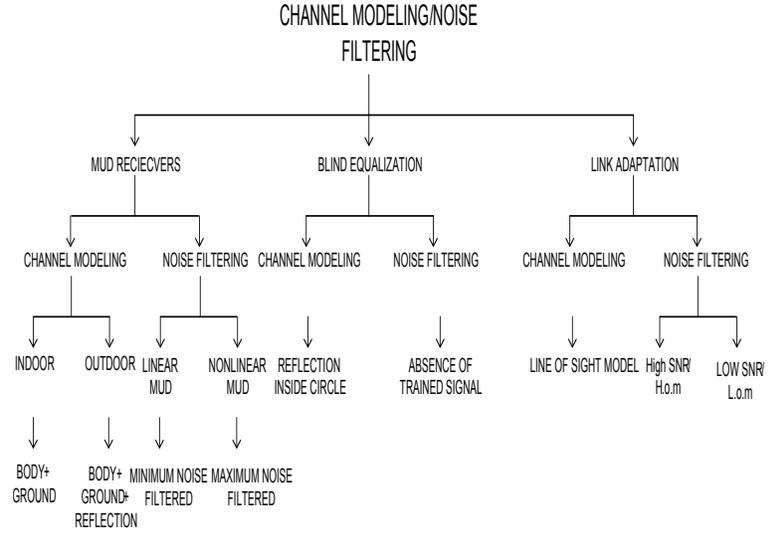}
  \caption{ Hierarchy for Channel Modeling and Noise Filtering }
\end{figure}

\section{Data Broadcast in IEEE 802.15.4 and ZigBee}

Energy consumption is an important issue in wireless sensor network and same is the case associated with WBAN. Multiple sensor nodes are working simultaneously for acquisition of useful information from the patients in WBAN. The key idea in this regard is to use IEEE 802.15.4 standard along with ZigBee network which is considered to be a low data rate and low cast infrastructure [4]. IEEE 802.15.4 defines two types of devices: Full Function Device (FFD) which can serve as a coordinator or a regular device and a Reduced Function Device (RFD) which is a simple device that associate and communicate only with an FFD. Based on IEEE802.15.4, ZigBee specifies the standards for network and application. ZigBee network layer is responsible for assigning addresses and builds a hierarchical tree topology [4]. Network parameters such as maximum allowable number of children's, $n_{chl}$ and maximum level of logical tree, $d_l$, are managed by a coordinator. A coordinator acts as a root of the tree with address zero.

On-tree self pruning broadcast algorithm keeps the energy efficient broadcast in account for ZigBee network and decides whether to rebroadcast or not after receiving a packet.  When a source node broadcasts a packet, then all its 1-hope neighbors receives this packet. This may cause collision
when multiple nodes are communicating and accessing the channel at the same time. To avoid such collisions, each node waits for random period of
time before rebroadcasting. During this waiting state, forward node, $'x'$ only needs to cover 1-hope distance i.e., $N(x)-N(Y)$. If node $x$ learns that all its 1-hope neighbors are already covered before time out then it does not need to rebroadcast [4].  This self pruning algorithm can perform poorly when applied to zigBee networks because of unavailibilty of 2-hope neighbor's information.

An optimal On-tree forward node Selection algorithm (OOS) resolves the problem and is assumed to guarantee convergence of the whole network [4]. This algorithm tries to reduce the size of the candidate forward node set $'S'$ and to be covered set $'C'$. If node $x$ is the source or forward node, it will rebroadcast and cover all its 1-hope neighbors, therefore, it does not need to be selected as a forward node again, as well as, all its neighbors need not to be covered again.


 \begin{figure}[ht]
  \centering
   \subfigure[]{\includegraphics[scale=0.35]{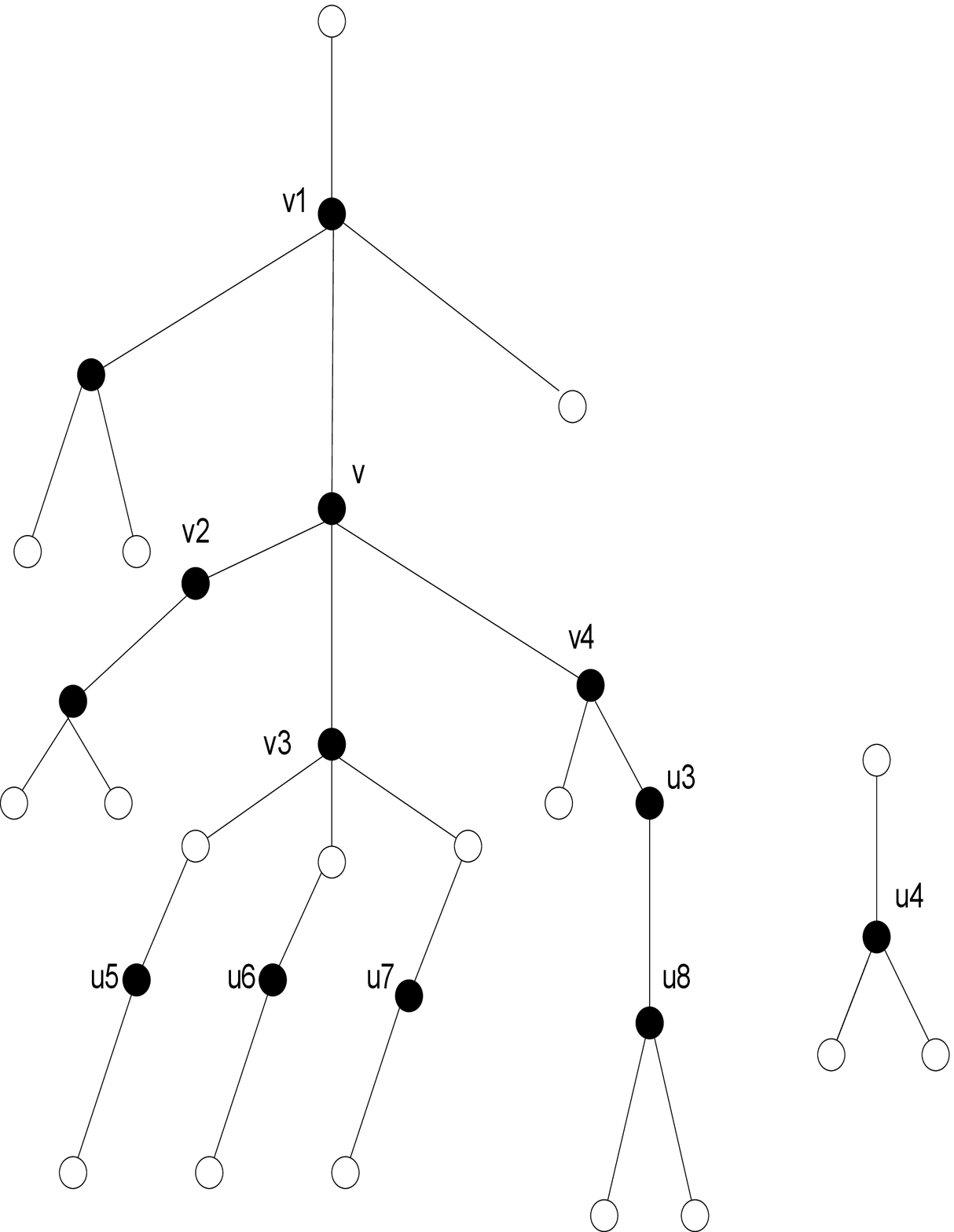}}
   \subfigure[]{\includegraphics[scale=0.35]{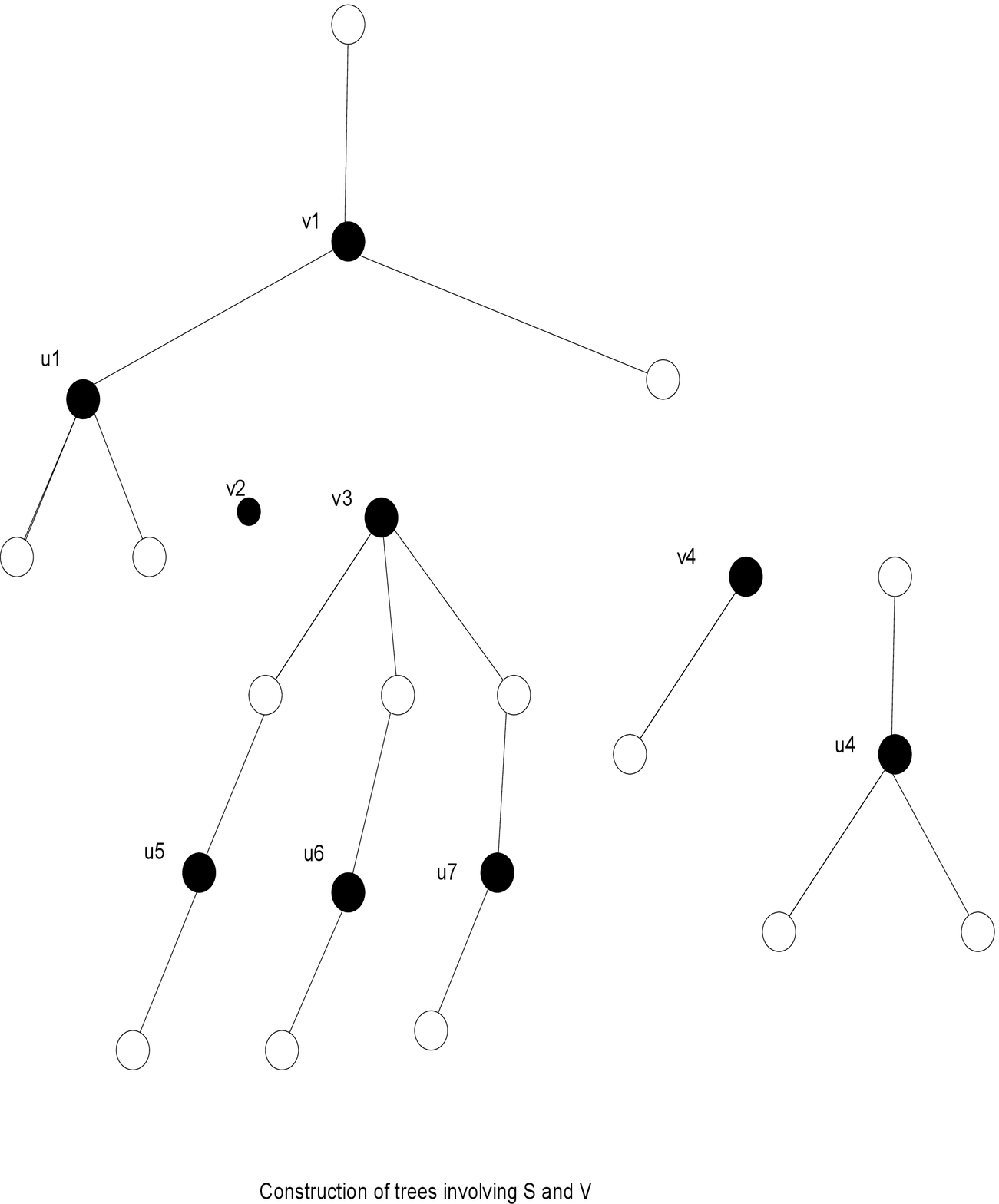}}
   \caption{OOS Algorithm in Local ZigBee using Tree Topology}
\end{figure}

Fig. $6a$ shows ZigBee tree topology with node $v$ at the centre and four tree neighbors v1, v2, v3 and v4. u1....u8 are other eight 1- hope
neighbors which can be located anywhere on the logical ZigBee tree. Given the network addresses of v1 to v4 and u1 to u8. Node V can further
identify their parent and children.

Fig. $6b$ shows principle OOS which describes set size of the candidate forward node S and C. OOS algorithm goes through level by level in a top-to-bottom direction and left to right at each level.

%

\section{Energy Utilization}
In this section, we discuss and compare energy utilization in selected three techniques; 1) MUD receivers, 2) Novel Link Adaptation, and 3) Blind
Equalization for modeling channels. WWBANs use a central device (i.e., hand held computer) and multiple sensors, simultaneously for communication. Multi user interference is a phenomenon which leads in the degradation of the received signal and certain receivers based on RAKE do not cancel this effect [1]. In BAN environment Multi user based receiver is considered to be an appropriate solution for the BAN environment where several devices communicate simultaneously [1].

\begin{algorithm}[H]
\caption{Algorithm for MUD receivers}
\begin{algorithmic}[1]
\STATE $ Discrete \;equvalent \;of \;received \;waveform \leftarrow \;r_1$
\STATE $ Frames \;samples \leftarrow \;N_s$
\STATE $ Mean \;square \; error \leftarrow \;MSE$
\STATE $ Filter \;tapes \leftarrow \;W$
 \STATE $Received \;output \leftarrow \;R_{out}$
 \IF{Choice = 1}
 \STATE $ Use \;linear \; MUD $
 \STATE $r_1=\sum_{\mu=1}^{N\mu}\sum_{k=-\infty}^{\infty}\alpha_{k}^{\mu}\rho^{\mu}(l-kNs)+n1$
 \STATE $ Calculate \;'N_s' \; from \; 'r_1' $
 \STATE ${w}=[w_1....w_n]$
 \STATE $ Choose \;'W' \; such, \;that \;minimizes \; 'MSE' \; $
 \STATE $ R_{out} = \;ISI + \; MUI +\; Noise  $
\ELSE
\STATE $ Use \;non-linear \;MUD$
\STATE $ For \;Feed \;Forward \;Filter \;(FFF)$
\STATE $ Rpeat \;steps \;8-10 $
\STATE $ For \;Feed \;Back \;Filter \;(FBF)$
\STATE $ (N_s-(N_s-1)) \leftarrow \;Status$
\STATE $ R_{out}$ = FFF + FBF
\STATE $ R_{out}$=\;ISI + MUI
\ENDIF
 \end{algorithmic}
 \end{algorithm}

\subsection{Multiuser Detection Based Recievers}
A  signal consists of following components; a) desired signal energy spread over several multiple symbols, b) ISI due to channel memory, c)
MUI due to other users signals, and d) additive white Gaussian noise is received [1]. MUD based receiver is used to extract effective energy in the presence of all the components. To effectively extract energy from different multipaths in the presence of ISI, MUI and AWGN. MUD receiver is implemented in two phases; 1) Linear MUD receivers and 2) non-Linear MUD receivers [1].

\subsubsection{Linear MUD Recievers}
A linear MUD receiver uses Weiner-Hopf equation to minimize mean square error. The effects of ISI, MUI and noise are mitigated jointly, however,
receiver cannot distinguish between MUI, ISI and noise. To differentiate between noise and ISI, another approach is proposed which is known as Non-linear MUD receivers.

\subsubsection{Non-Linear MUD Recievers}
Estimation of previous symbol could be utilized while estimating the current symbol. This could be achieved by exploiting decision feedback
principle. A Non-Linear MUD receiver consists of two filters: one is Feed Forward Filter (FFF) and Feed Back Filter (FBF). FFF resembles Weiner combiner, and the FBF has at its inputs the sequence of decisions on previously detected symbols. FBF is intended to remove the part of the ISI from the current symbol which is caused by previously detected symbol. By minimizing mean square error through Linear and Non-Linear receivers, retransmission can be mitigated. Thus, minimizes energy consumption.

\subsection{Novel Link Adaptation}
Novel Link Adaption (LA) mechanism improves performance and achieves significant result while reporting measured data to a coordinator with different modulation schemes [2]. A variable data rate scheme is proposed to reduce the average power consumption [2]. This scheme considers channel conditions which would work in a better way. The usage of  lower bit rates (low SNR/SIR) and higher bit rates to reduce probability of error is well solved in [2]. The use of dynamic modulation/coding schemes (O-PSK, L-DSPK) by keeping channel response in view can result
beneficially in reduction of interferences. Thus, minimizes energy consumption through avoiding re-transmissions.

\subsection{Blind Equalization Algorithm}
Conventional equalization techniques rely on the transmission of training signals. This relay leads to a reduction of channel bandwidth and sources
allocation. Constant Modulus Algorithm (CMA) is suitable for blind wireless channel equalizer, because of its robustness over the violation of
perfect blind equalization condition gives full bandwidth utilization [3]. Transmission time is always an important factor in terms of cost and
transmission power being utilized. Large data rate can be achieved through lossless compression algorithm. In wireless communication systems, most
of the energy is consumed for transmission. Multiple interferences lead to the effect of retransmission during signal transmissions. In this
study different techniques are discussed which addresses the minimization of different interferences/disturbances.
\section{Impairments}

Simultaneous communication among sensors and central devices (hand held computers) yields multiple interferences like ISI, MUI, and noise. RAKE
based receivers do not cancel these effects, therefore, MUD receivers are used. Whereas, determination of the coefficients of the combiner is carried through training sequence method.

\subsection{Linear and Non-Linear Equalizers }
By using MMSE criterion to determination of tap coefficients is given in [1] as:

\begin{eqnarray}
\frac {Tr}{Tn}=Ns>1
\end{eqnarray}

Moreover, considering the above condition for sampling rate in [1] as:

\begin{eqnarray}
r_1=\sum_{\mu=1}^{N\mu}\sum_{k=-\infty}^{\infty}\alpha_{k}^{\mu}\rho^{\mu}(l-kNs)+n1
\end{eqnarray}

For   u =1

\begin{eqnarray}
r_1=\sum_{k=-\infty}^{\infty}\alpha_{k}^1\rho(l-kNs)+\sum_{\mu\neq1}\sum_{k=\infty}^{\infty}\alpha_{k}^{\mu}\rho^{\mu}(l-kNs)+n1
\end{eqnarray}

 Representing the MUI as $m1$ given in [1] as:

 \begin{eqnarray}
 r-1=\sum_{k=-\infty}^{\infty}\alpha_{k}^{1}\rho(l-kNs)+m1+n1
 \end{eqnarray}

Filter tapes ${w}=[w_1....w_n]$ minimize MMSE $E(\|a_k^1-Wr_k\|^2))$ are desired to be chosen, and are obtained through Weiner-Hopf equation [1]. $w=\gamma_ar\Gamma_{rr}^-1$ where, $\Gamma_{rr}=E[r_kr_k^T]$ denotes received vector auto-correlation matrix and $\Gamma_{rr}=E[a_k^1r_k^T]$ represents cross correlation between known transmitted symbol and the corresponding received samples. The receiver tries to mitigate the effect of MUI, ISI and noise jointly. However, it does not distinguishes MUI and ISI from noise, thus, the output still contains some amount noise. In order to mitigate the effect of ISI, a Non-Linear filtering approach based on FFF and FBF is used. FBF  has at inputs the sequence of decisions on previously detected symbols. The main aim of this decision is to remove ISI from the current symbol caused by previous symbols.

MMSE criterion is applied to optimize the coefficients of the filters. It is essential to note that input samples to FFF are spaced $Tr$ seconds
apart while input samples of the FBF are spaced $Ts$ seconds apart. The equalizer output is represented in [1] is given as
$x_k=w_{ff}r_k+w_{fb}$ where, the row vector $wff$ denotes $Ns$ length of FFF and $wfb$ is the $Nb$ length vector representing FBF. The set
of past decisions is represented by $a_k^1=[a_k^1,....a_{k-Nb}^1]^T$ as given in [1].

\begin{figure}[!t]
 \subfigure{\includegraphics[height=4  cm,width=6 cm]{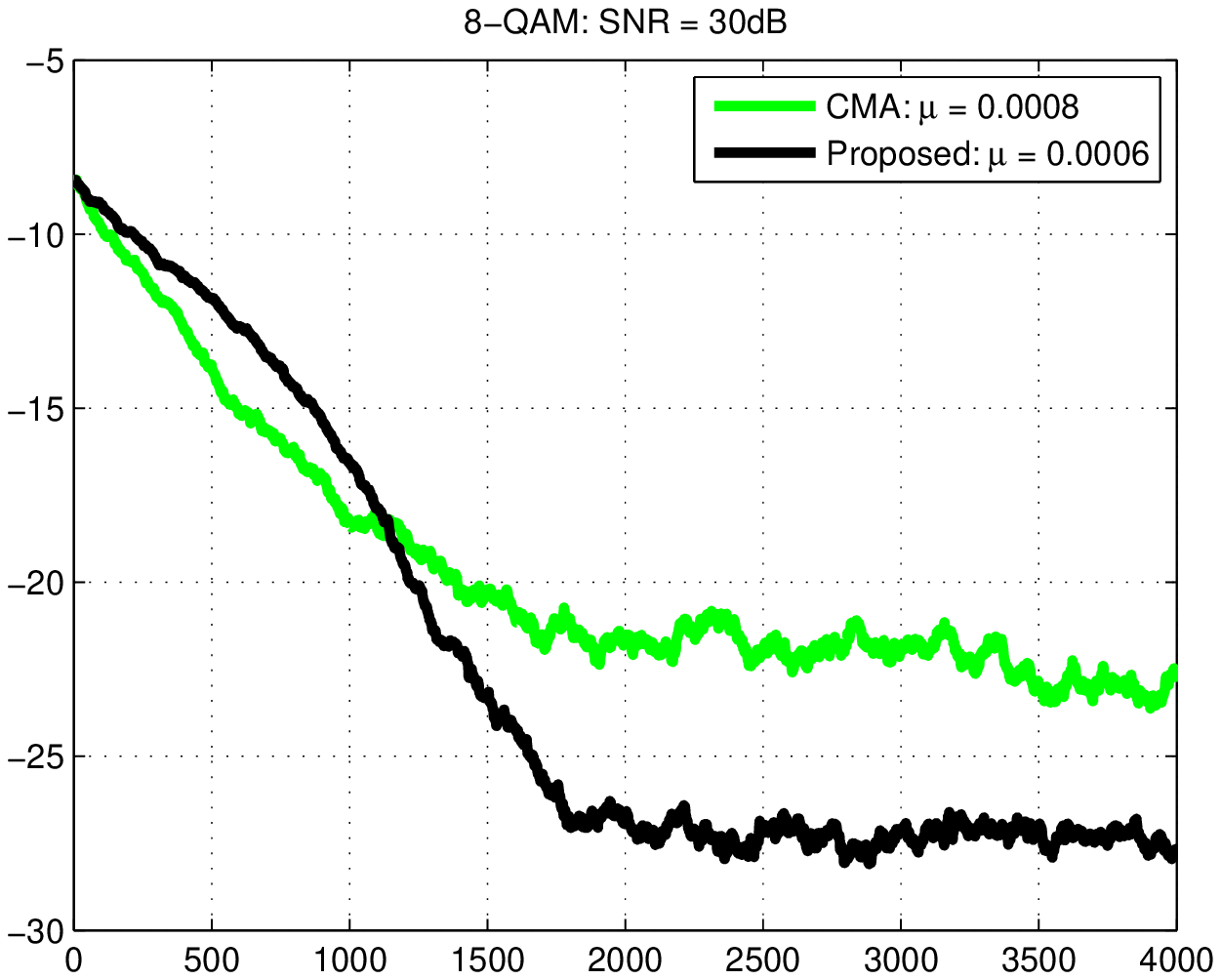}}
   \vspace{-1cm}
 \subfigure{\includegraphics[height=4 cm,width=6 cm]{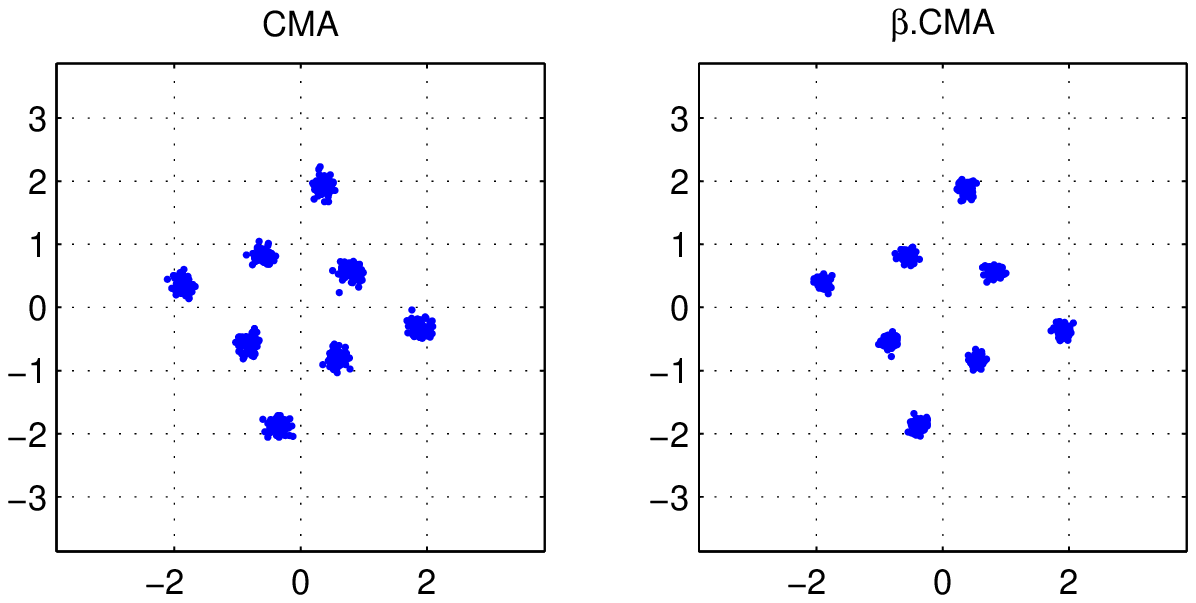}}
   \vspace{-0.6cm}
  \subfigure{\includegraphics[height=4 cm,width= 6 cm]{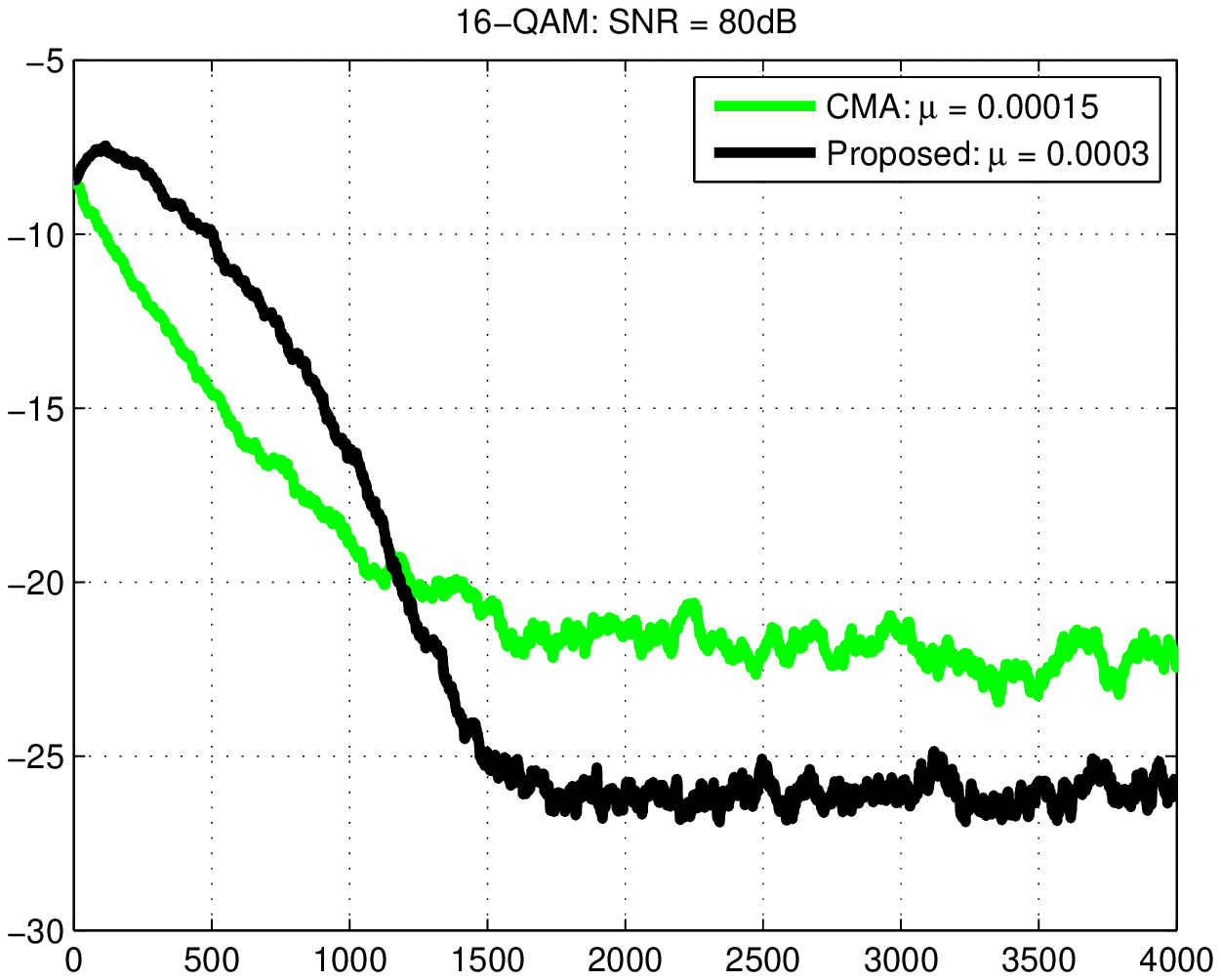}}
   \subfigure{\includegraphics[height=4  cm,width=6 cm]{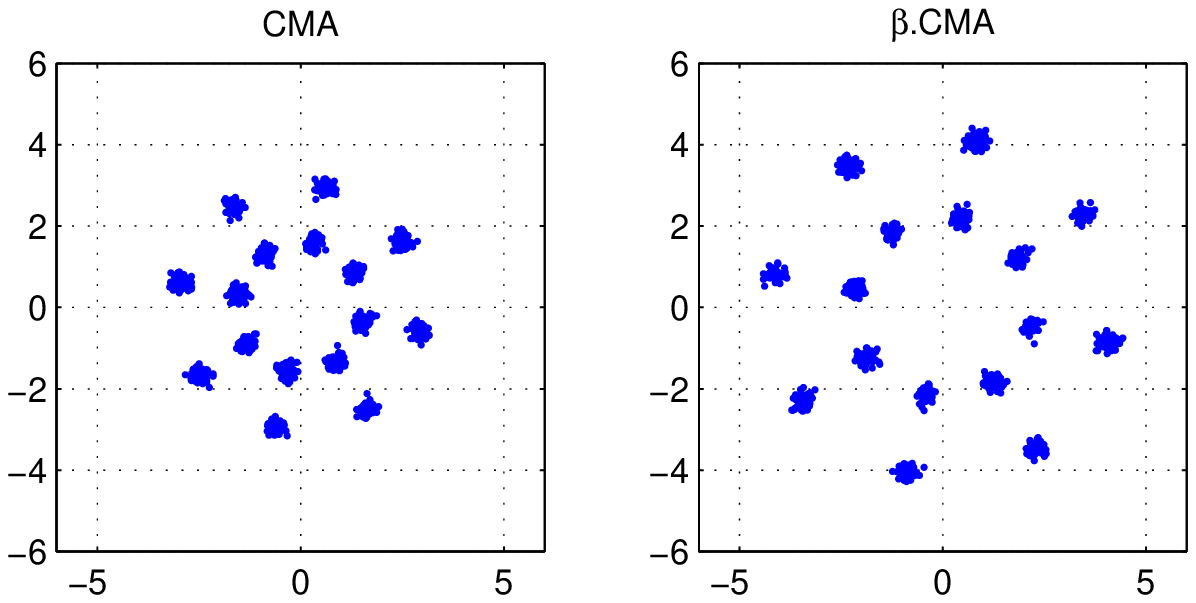}}
   \vspace{-1cm}
  \caption{Blind Equalization Using CMA, 8-QAM and 16-QAM }
\end{figure}

\subsection{Novel Link Adaptation }
Another optimized proposed solution for the removal of ISI is use of dynamic modulation/coding schemes according to channel response as given in
[2]. Considering 3GPP proposal for WBANs, the channel loss can be calculated from [2] as:

\begin{eqnarray}
A(d)[dB]=A_o+10nlog(d/d_o)+s
\end{eqnarray}

where $d$ is the distance between transmitter and receiver, $A_0$ is the path loss in dB at a reference distance; d0 (A0 = 35.2 dB for d0 = 0.1 m),
$n$ is path loss exponent (n = 3.11), and $s$ is a random Gaussian variable, having zero mean and standard deviation $\sigma =6.1 dB.$ for proper
reception of packet, we assume that following two conditions should be achieved:  a) $PR >PR_{min}$, where PR is the received power given
by: $P_R[dBm]=P_{tx}[dBm]-A[dB]$, $P_{tx}$ and $P_{Rmin}$ are transmit power and receiver sensitivity, respectively [2]. Total power received from
interfacing nodes is denoted by $I$, whereas, $C$ represents the power received from useful transmitter. $C/I\mid_{min}$ is the minimum SIR ensuring the correct reception of a packet. These values are function of modulation schemes being used.

\subsection{Blind Equalization Algorithm }
A bandwidth limited channel with high data rates is always effected by ISI. This is due to adjacent symbols on the output of the channel which
overlap each other and causing degradation of error performance. An equalization filter attempts to extract the transmitted symbol sequence by
de-convolving them with the inverse version of channel. Meanwhile, multiplexing techniques emerge to be a powerful source for minimizing
system bit error rate.

Cyclic prefix (longer then channel length) within OFDM systems by converting linear convolution into cyclic one mitigates the effect of ISI
significantly [3]. To utilize full bandwidth of the channel adaptive blind equalization systems are introduced in order to minimize the effect of
ISI [3]. Transmission of training signals in conventional equalization techniques leads to reduction in bandwidth. Whereas blind equalization do not rely on the training sequence, hence, results in a better consumption of allocated resources.

Because of robustness over violation of Perfect Blind Equalization (PBE) conditions, CMA is an attractive choice for the
researchers. Dithered Signed-Error Constant Modulus Algorithm (DSE-CMA ) changes the input signal to the equalizer by a non-subtractive
sinusoidally-distributed signal before a sign operation is applied [3]. The output of the equalizer using multi-rate model from [3] is given as:

\begin{align*}
y(n)&=r^H(n)f(n)+W^H(n)f(n)\\
&=X^H(n)Hf(n)+W^H(n)f(n)
 \end{align*}

where, $X=[x(n),x(n-1),....,x(n-N_x+1)]^t$ denotes a finite-length symbol vector. The length of this symbol vector is given by $N_x=(N_h+N_f-1)/2$,
$f$ represents a column vector of fractionally-spaced equalizer coefficients with a length $N_f$ . The column vectors $r(n)$ and $w(n)$ represent the time decimated $N_f$ received samples and white Gaussian noise, respectively. A Hermitian operator is denoted by (.)H, and matrix transposition is symbolized by (.)T. Matrix H symbolizes a $N_x*N_f$ time-decimated channel convolution of the GBHDS channel model.

Signals when pass through a channel undergo various forms of distortions. Most common is ISI. Most of the times, in practical systems, channel characteristics are not known as prior. Therefore, adaptive equalizers are used. Blind equalization as an adaptive equalizer can compensate amplitude and delay distortion of a communication channel only by using channel output samples and knowledge of basic statistical properties of the data symbol. We compared simulated results for blind equalization using CMA. We used picchiprati channel model for our experiments. The transmitted signals are modulated with 8-QAM and 16-QAM, respectively. From the fig. 8 given it is clear that step size $\mu=0.0006$ and $\mu=0.0003$ for 8-QAM and 16-QAM, respectively are sufficiently enough to minimize MSE at steady-state.Thus, convergence is achieved.

\section{Conclusion}
In this paper, we addressed some of the problems that a body area network still has to deal with for efficient performance in order to provide an optimized solution for the cure of elderly people. We investigated some of the issues and tried to sort out their possible solutions. Channel
modeling, effect of interferences/disturbances, energy consumption, and noise filtering are some of the issues which should be given importance on
the equal basis to attain better performance. We also discussed some simulated results describing the use of different modulation/coding schemes.
Decision is taken on the basis of channel characteristics.

\end{document}